\documentclass[%
superscriptaddress,
preprint,
 amsmath,amssymb,
 aps,
prb,
]{revtex4-2}

\usepackage{graphicx}
\usepackage{multirow}
\usepackage{dcolumn}
\usepackage{bm}
\usepackage{xcolor} %
\usepackage{siunitx}
\usepackage{mathtools}
\usepackage{hyperref}
\usepackage{tikz}
\usepackage{subcaption}

\usepackage[mathlines]{lineno}

\def\njust{Department of Applied Physics, Nanjing University of Science and Technology, Nanjing 210094, China}

\begin{document}

\title{Pressure dependence of liquid iron viscosity from machine-learning molecular dynamics}

\author{Kai Luo}
\email{kluo@njust.edu.cn}
\affiliation{\njust}%


\author{Xuyang Long}
\affiliation{\njust}%



\author{R. E. Cohen}
\email{rcohen@carnegiescience.edu}
\affiliation{Earth and Planets Laboratory, Carnegie Institution for Science,
5241 Broad Branch Road, NW, Washington, DC 20015, USA}%

\date{\today}
\begin{abstract}
We have developed a machine-learning potential that accurately models 
the behavior of iron under the conditions of Earth's core. By performing numerous nanosecond scale equilibrium molecular dynamics simulations, the viscosities of liquid iron
 for the whole outer core conditions are obtained with much less uncertainty. We find that the Einstein-Stokes relation is not accurate for outer core conditions. 
 The viscosity is on the order of 10s \si{mPa.s}, in agreement with previous first-principles results.
We present a viscosity map as a function of pressure and temperature for liquid iron useful for geophysical modeling.
\end{abstract}

\maketitle

\section{Introduction}
\label{sec:introduction}
The viscosity of liquid iron is an important property for modeling the formation and dynamics of planetary cores and the generation of planetary magnetic fields \cite{Roberts2000}. Iron viscosity has been studied at Earth core conditions using first-principles based molecular dynamics \cite{Vocadlo2003,Alfe2010,Fomin2013,Xian2019}
and there is general consensus on values of about 10 \si{mPa.s}. However, there is a long-standing problem  \cite{Secco1995} of a 13-order of magnitude difference between mineral physics and seismological estimates of the viscosity. This seemed bolstered by Ref. \cite{Smylie2009}, but was rebutted by \cite{Zharkov2009}. Cormier \cite{Cormier2009,Cormier2022} suggests a glassy layer at the base of the outer core with viscosity of $10^9$ \si{Pa.s}, 11 orders of magnitude higher than the values from molecular dynamics or extrapolations of lower pressure experimental data\cite{Mineev2010,Leblanc1996,Rutter2002,Dobson2000}. Thus further investigations of viscosity in iron are warranted. 

Seismological observations and 
mineral physics studies have constrained that Earth's core is 
mainly composed of iron (Fe) alloyed with other light elements such as nickel (Ni), silicon (Si), oxygen (O), carbon (C), 
and hydrogen (H) \cite{McDonough1995}. 
Iron is the dominant component and the most crucial element for establishing a baseline
understanding of core materials. Liquid iron's transport properties, 
such as thermal conductivity, magnetic diffusivity, and shear viscosity, are particularly relevant for dynamo simulations under Earth's core conditions. 
For example, in the Boussinesq formulation of the dynamo theory, many dimensionless parameters  such as the Prandtl number, 
the magnetic Prandtl number, the Ekman number, 
and the Rayleigh number \cite{Boussinesq} depend on the viscous behavior of the liquid. 
 
The Green-Kubo formula derived from the fluctuation-dissipation theorem is a fundamental and in theory exact way to compute viscosity from molecular dynamics simulations  \cite{Allen2017}. These results can be compared with viscosities obtained using the Stokes-Einstein relationship between diffusivity and viscosity \cite{deWijs1998}, which requires an estimate of the atomic size and the boundary conditions for the sliding of the particle through the fluid. High pressure experiments, on the other hand, are usually based on Stokes-Einstein, with falling or rising spheres driven by density difference, but which typically allow only determination of viscosities along the melting line, and have not been possible at planetary core conditions. 

First-principles molecular dynamics (FPMD) provides accurate estimates of viscosity\cite{Malosso2022, Silvestrelli2023}, but   
it is computationally expensive and difficult to systematically study size effects \cite{deWijs1998, Alfe2000, Pozzo2013, Li2021, Li2022_PEPI, Zhu2022, Cong2025}. 
Molecular dynamics with empirical potentials, such as the Sutton-Chen embedded atom model (EAM), 
has been generated to overcome this limitation \cite{Desgranges2007,Xian2019}, but the accuracy at core conditions is unknown. Advances in simulation techniques and capabilities allows for the generation of 
effective machine-learning many-body potentials with first-principles accuracy. %
Recently, machine-learning potentials have been used for studying elasticity, viscosity, and melting temperature of hcp iron 
and superionic FeH$_x$ at Earth's inner core conditions \cite{Li2022GRL,Yuan2023, Wu2024, Xu2025}.

In this work, adopting the deep potential \cite{DeepMD2018} in large-scale simulations
and using direct calculations in the Green-Kubo formula, 
we study the variation in diffusivity and viscosity  for liquid iron as functions of pressure and temperature.

\section{Methods}
\label{sec:comp_methods}

\subsection{Potential Training}
\label{sub:ml_potential}
We use {\sc DPGEN} \cite{DPGEN2020} to develop a many-body potential capable of accurately describing the thermodynamic conditions of Earth's outer core. 
DPGEN is a concurrent learning platform designed for exploration in complex spaces \cite{Zhang2021,Yang2021,Chen2023,Wu2023} and has a workflow involving an iterative cycle of three stages: exploration, labeling, and training.

To obtain the machine-learning (ML) potential, we start with  FPMD of ideal {\it hcp} iron with $c/a=\sqrt(8/3)$ using the Vienna ab initio Simulation Package (VASP) \cite{VASP}. We consider densities of $\rho=10.0, 11.0, 12.0, 13.0$ \si{g/cm^3} 
and temperatures $T=4000, 5000, 6000$ \si{K}. 
Electrons at elevated temperatures are populated according to the Fermi-Dirac distribution density functional theory \cite{Mermin1965}.
Electron-ion interactions are described via the VASP
projector-augmented-wave (PAW) potential with $14$ valence electrons, which shows pressure error less than 4 GPa for ICB conditions against all-electron LAPW tests\cite{Sun2018}. The Perdew-Burke-Erzernhof exchange-correlation 
density functional is used throughout \cite{KohnSham1965,PBE1996,PBE1996Erratum}.
 Orbitals are expanded in the planewave basis with the energy 
cutoff of $950$ \si{\eV}, which is tested to yield converged energies with error lower than $0.5$ \si{\meV} per atom.
The ionic temperature is controlled in the $NVT$ ensemble and time step of $0.2$ \si{\fs} during training is chosen to ensure negligible energy drift. With $128$ atoms in the simulation cell, only $\Gamma$ point sampling is used for the DFT computations. We run the simulation for a duration of $3$ \si{\ps}, using a maximum of 1130 bands to ensure the highest occupied band had an occupation of less than $10^{-6}$.

We initialize an ensemble of 4 models by selecting frames from FPMD trajectories every 50 steps. 
During the exploration stage, the models traverse conditions ranging from $100$ \si{GPa} to $360$ \si{GPa} and from $3000$ \si{K} to $7000$\si{K}.
This stage stops until the accuracy ratio reaches 100\%. 

The potential is constructed by decomposing the energy into atomic fragments and introducing covariant descriptors 
through the use of local coordinates. Atomic energies and forces are computed using only neighbors within a specified
 cutoff radius.
For the embedding net, we use a three-layer neural network with neurons $[25, 50, 100]$, and for the fitting net,  a network size of $[240, 240, 240]$ is chosen.
We set a two-atom embedding descriptor \textit{``se\_e2\_a"} with a cutoff radius of $5.5$ \si{\angstrom}, which encodes both angular and radial information of neighboring atoms. 
To effectively select configurations for labeling, we choose $54$ atoms ({\it bcc} and {\it hcp}) frames 
with model deviations in the range of $0.30$ to $0.50$. 
We adopt the same setting as that of FPMD for labeling, 
except with a denser k-point density of $0.5$ for the Brillouin sampling.  The number of bands $570$ is sufficient for this system size. To account for electronic free energy, we employ a deep potential that is dependent on the electron temperature \cite{Zhang2020, Wu2024}. 

After collecting all the frames explored, we conduct a long production training for $2,000,000$ steps. 
We apply the compression technique\cite{DPCompress2022} with a cutoff radius of $5.0$ \si{\angstrom} for speedup without losing accuracy, which yields converged
radial distribution functions compared to the FPMD results. The trained potential exhibits root-mean square error (RMSE) of $12$ \si{meV} 
for the energy per atom and $301$ \si{meV/\angstrom} for the force on the testing dataset of over 550 frames.

\begin{figure}[htbp]
    \centering
\includegraphics[width=0.95\textwidth]{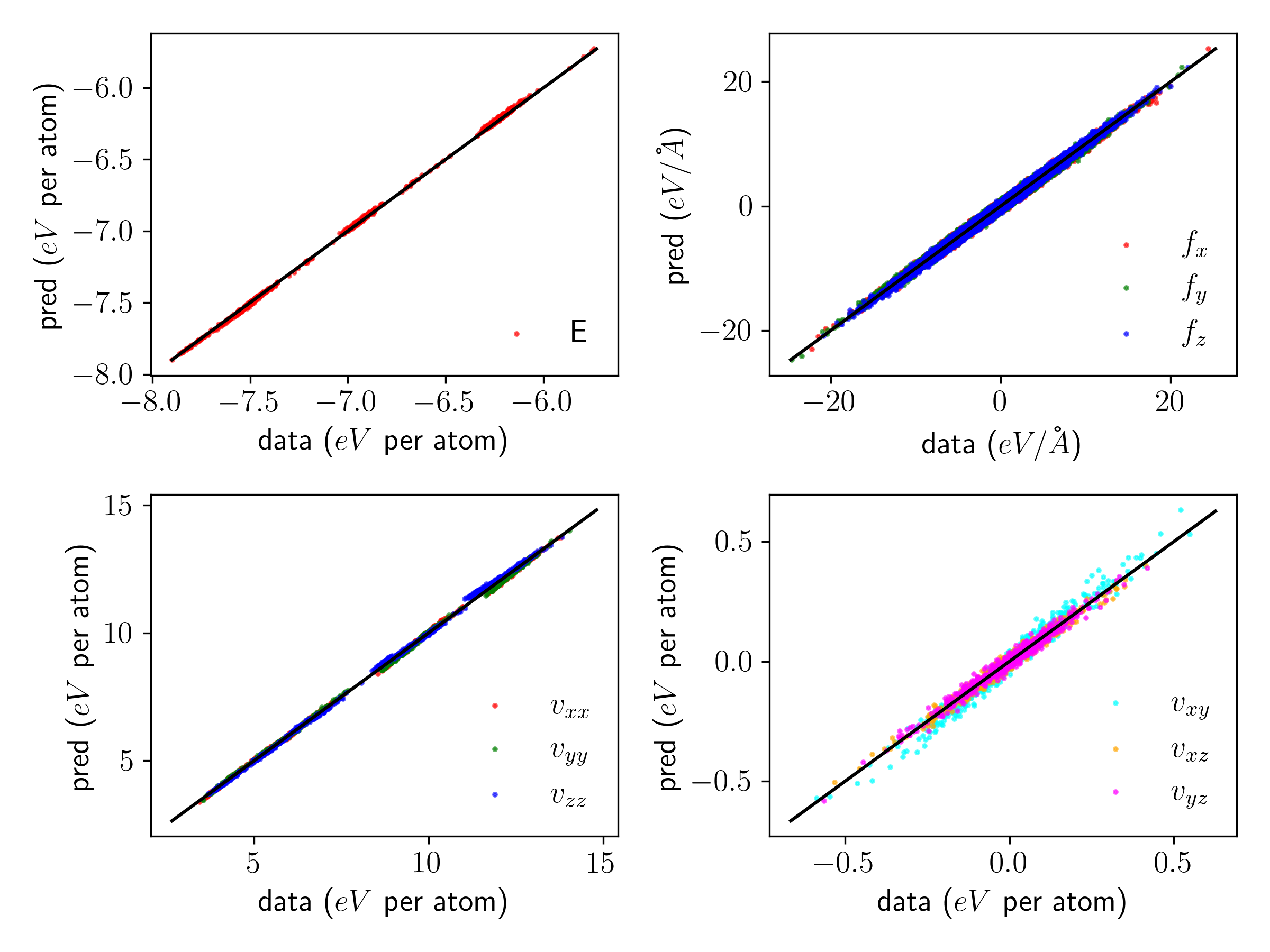}
\caption{
Predictions of energy, forces, and virials 
from the machine-learning potential \textit{versus} that from the first-principles calculations. The virials, $\sum_i \vec{R}_i \vec{F}_i$, are the potential component of the stress.
Top left panel for the energy, top right panel for the forces in 3 directions, bottom left panel for the diagonal virials,
bottom right for the off-diagonal virials. }
\label{fig:error}

\end{figure}

For the testing set, we plot the predicted energy/forces/virials of the machine learning potential against the first-principles forces (Fig. \ref{fig:error}). 
The energy and force agreement is satisfactory. The diagonal virials agree well, while the off-diagonal components are slightly worse which is generally acceptable at such high temperatures. 

\begin{figure}[htbp]
    \centering
    \includegraphics[width=0.95\textwidth]{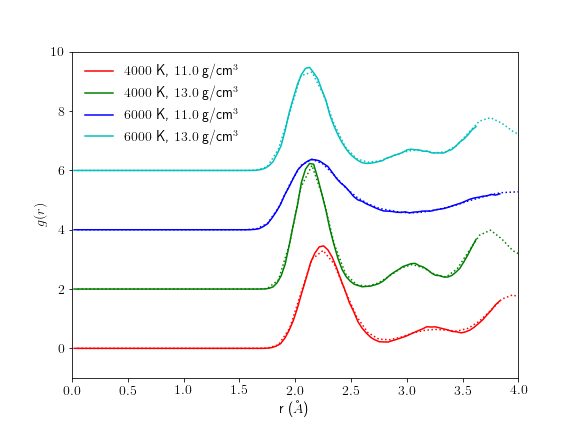}
\caption{The radial distribution function (RDF) is calculated for a few representative Earth's core conditions 
to validate the accuracy of our trained deep potential molecular dynamics (DPMD) model. 
The RDF curves obtained from our DPMD simulations are shown as dotted lines, 
and agree well with those obtained from first-principles molecular dynamics (FPMD) simulations, 
shown as solid lines. To aid visualization, shifts have been applied to the DPMD curves. 
    }
    \label{fig:rdf}
\end{figure}

To validate the potential, we compare the radial distribution function (RDF) against FPMD predictions of a few 
 representative conditions, as shown in Fig.(\ref{fig:rdf}). Agreement is excellent.

\subsection{Viscosity Calculation}
\label{sub:viscosity_calc}

We use the LAMMPS package \cite{LAMMPS} with a modified pair style  to conduct
deep potential molecular dynamics (DPMD) simulations.
To investigate the size effects and ensure the convergence of our results, we perform a viscosity convergence study using the Green-Kubo formula (Eq. 1)
with equilibrium molecular dynamics trajectories. The viscosity 
 $\eta(\tau)$ is computed by integrating the off-diagonal stress tensor $P_{\alpha\beta}$'s autocorrelation function 
 $\left\langle P_{\alpha\beta}(0) P_{\alpha\beta}(t) \right \rangle$ from 0 to $\tau$,
 \begin{equation}
    \label{eq:gk}
    \eta(\tau)= \frac{V}{k_B T} \int_{0}^{\tau} \left\langle  P_{\alpha\beta}(0) P_{\alpha\beta}(t) \right \rangle dt
\end{equation}
 where $\alpha\beta$ takes $xy, yz, zx$, 
 and $V, k_B, T$ are the cell volume, the Boltzmann constant, and the temperature, respectively. 
 The final viscosity $\eta$ is taken when $\tau \to \infty$. 

\subsubsection{Parameters in the correlation function}
\begin{figure*}[htbp]
\centering
\begin{tikzpicture}{scale=0.9}
    \def\td{15} 
    \def\tw{6} 
    \def\ts{3} 
    
    \draw[thick, fill=gray!20] (0,0) rectangle (\td,0.5);

    \foreach \i in {0,1,3} {
        \draw[shift={(\td+1, -\i-1)}, thick, fill=gray!20] (\i*\ts-\td-1,0) rectangle (\tw+\i*\ts-\td-1,0.5);
    }
    \def\h{-2.5}
    \draw[black] (\tw/2+2*\ts,\h) circle (0.1cm);
    \draw[black] (\tw/2+2*\ts,\h-0.3) circle (0.1cm);
    \draw[black] (\tw/2+2*\ts,\h-0.6) circle (0.1cm);
    
    \draw[<->] (0,1.0) -- (\td,1.0);
    \draw (\td/2,1.0) node[above] {$t_D$};

    \draw[<->] (0,-2.0 + 0.25 ) -- (\ts,-2 + 0.25);
    \draw (\ts/2,-2.0 + 0.25) node[above] {$t_S$};

    \draw (\ts+\tw/2,-2.5+0.25) node[below] {$t_W$};
    \end{tikzpicture}
    \caption{A diagram depicting how the time intervals
     used in computing the shear viscosity from MD simulation data using the Green-Kubo formulation are related.
     The total length of the tape is $t_D$, with each correlation window of length $t_W$ and separated by $t_S$.}
     \label{fig:diagram}
\end{figure*}
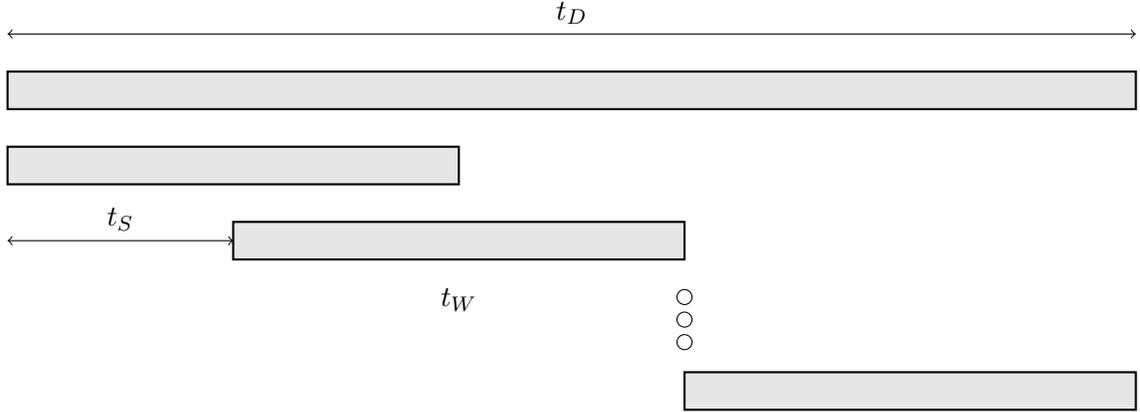

 A schematic diagram 
for computing the stress-tensor autocorrelation function is shown in Fig. \ref{fig:diagram}. The stress tensor is dumped every 5 time steps. With time step 1 \si{fs}, we set the total duration $t_D= 400000$. Depending on the 
actual simulation, the running viscosity might take longer to saturate. In this case, the upper limit of integration 
should be increased. Once the upper limit $t_c$ is set, the correlation width $t_W$ is typically 8-10 times of $t_c$.
The separation of origins $t_S$ does not affect the results when each origin of a window is statistically uncorrelated with other windows. 
Normally, the choice of $t_S = 300$ or greater suffices. In most cases, 
the setting of $t_c = 100, t_S = 1000, t_W = 1000$ is used.  Those cases close to the liquid-solid boundary require more careful tuning. 

\begin{figure}[htbp]
    \centering
\includegraphics[width=0.8\textwidth]{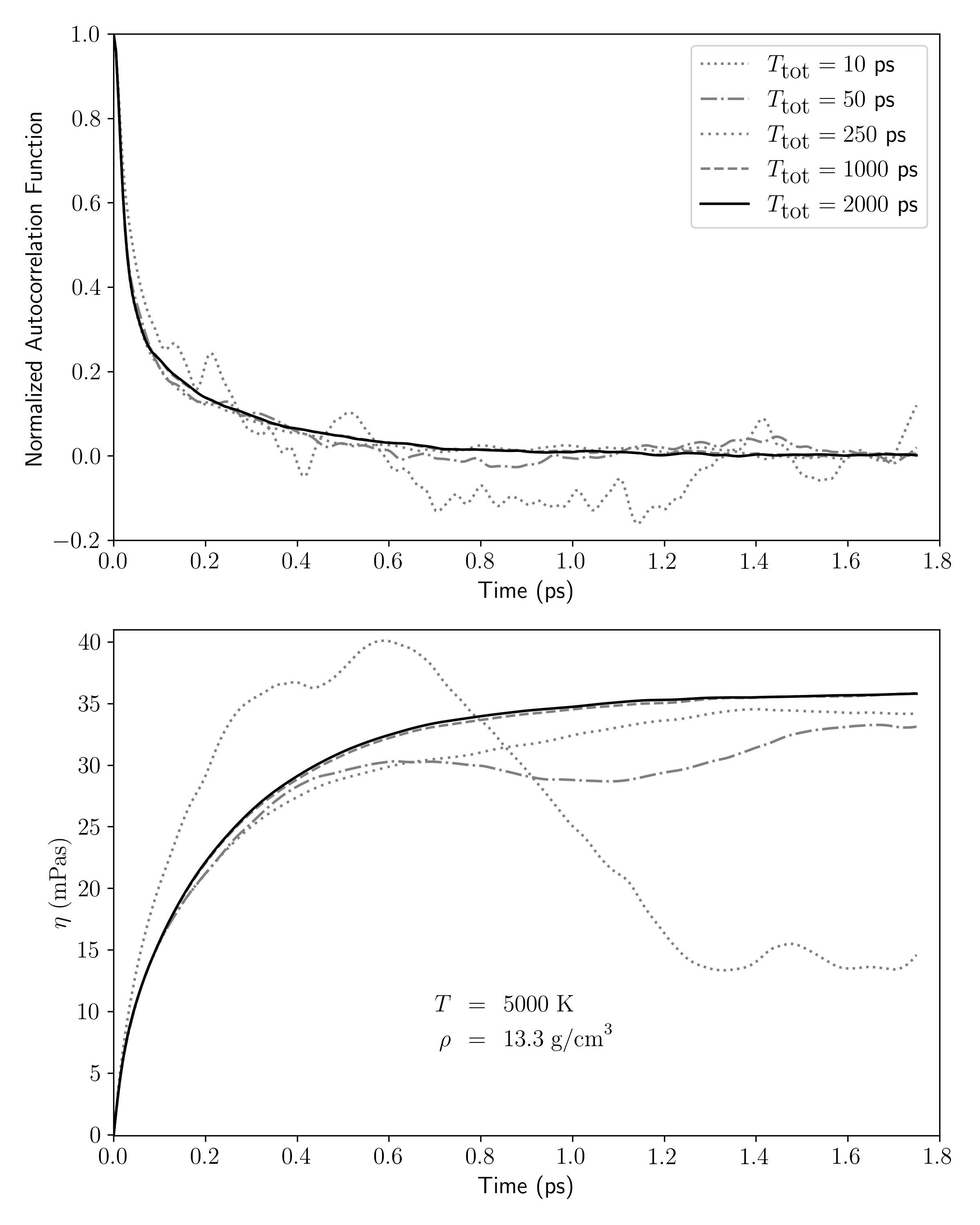}
\caption{Effects on the total duration of the simulation. Parameters are
$\omega = 5$, $\Delta t = 1$ \si{fs},  $t_W = 1000 $, $t_S = 1000$. Varying $t_D$ from 2000 up to 400000 shows that 1 \si{ns} significantly 
reduces the noisy tail of the correlation function.
}
\label{fig:time_effects}

\end{figure}

With dumping frequency $\omega = 5$, step size $\Delta t = 1$ \si{fs}, and correlation window size $t_W = 1000 $, 
the correlation function is averaged over a number of samples, 
 $N_{\textrm{orig}}$. Each sample lasts $\omega t_W \Delta t $ = 5 \si{ps}. With the total duration of $t_D$,
$N_{\textrm{orig}}$ can be estimated by an integer, $(t_D - t_W)/ t_S$.
To study the total time effects, by leveraging the efficiency of the machine-learning potential, we vary the total time of simulation $t_D$, we find that  
the choice of  $t_D \approx 400000$ effectively eliminates the fluctuation of the correlation function in the tail region (see Fig. \ref{fig:time_effects}). 

The initial structures is heated in an NVT ensemble 
to $1000$ K above the target temperature $T_{\textrm{target}}$ for $50$ \si{\ps} and 
is equilibrated at $T_{\textrm{target}}$ for another $400$ \si{\ps}. 
$\eta(t)$ tends to reach a plateau and then fluctuates around it.
To minimize the fluctuation, very long simulations are required and thus all production runs last $2$ \si{\ns}. 

To fit the asymptotic values, we employ a double exponential form \cite{Hess2002}, and the uncertainties are estimated using errors in the fitting coefficients. 

\subsubsection{Fitting function}
\begin{figure}[htbp]
    \centering
    \begin{subfigure}[b]{0.48\textwidth}
    \includegraphics[width=\textwidth]{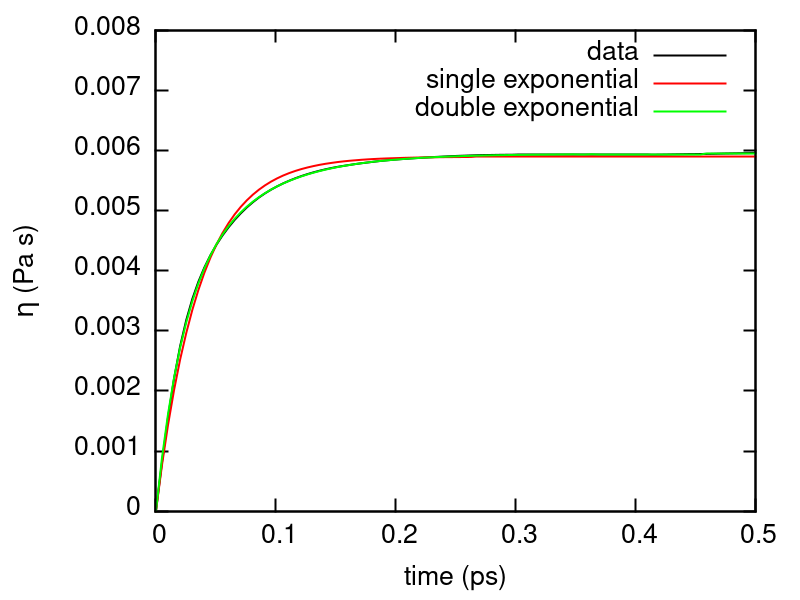}
    \caption{
    }
    \label{fig:two_fits}
    \end{subfigure}
    \hfill
    \begin{subfigure}[b]{0.48\textwidth}
    \includegraphics[width=\textwidth]{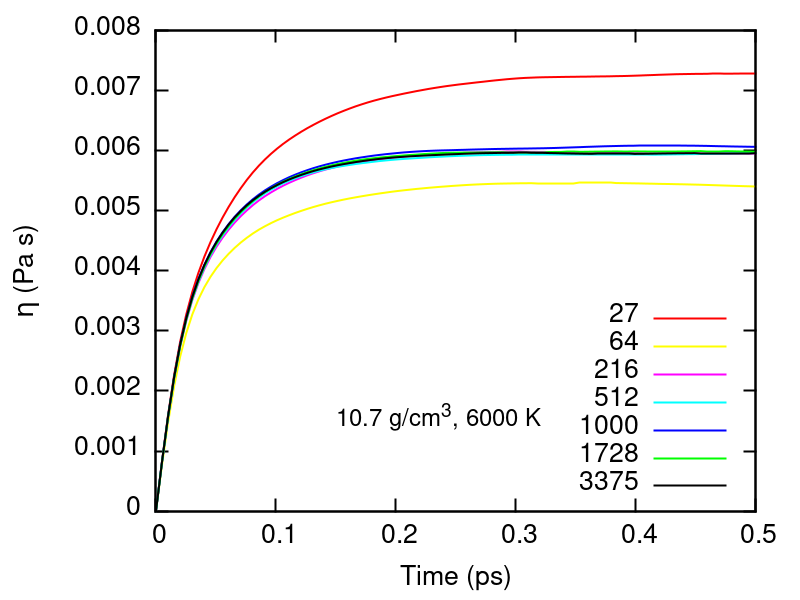}
    \caption{
    }
    \label{fig:size}
    \end{subfigure}
    \caption{(a) Running viscosity fits for two functions with the machine-learning potential. The condition of $\rho=10.7$ \si{g/cm^3}, $T=6000$ \si{K} is used
    (b) Running viscosity for various system sizes under the same condition. 
}
\label{fig:4}
\end{figure}

In determining the asymptotic behavior of $\eta(\tau)$, a fitting function is used. A simple exponential form 
\begin{equation}
    f(t) = A (1 - e^{-\frac{t}{\tau}})
\end{equation}
can gives an estimate of the viscosity. Taking $t\to\infty$, we obtain $A$ as the viscosity. 
A more accurate estimate is achieved using the double exponential form \cite{Hess2002}
\begin{equation}
    f(t) = A a \tau_1 (1 - e^{-\frac{t}{\tau_1}}) + A (1-a) \tau_2 (1 - e^{-\frac{t}{\tau_2}})
\end{equation}
and,  $\eta=Aa \tau_1 +  A (1-a) \tau_2$. The fitting is performed using the nonlinear least-squares Marquardt-Levenberg algorithm.
To choose which form better describes the viscosity, we compare two fitting functions on the calculated viscosity data.
The double exponential form is more accurate (Fig. \ref{fig:two_fits}). 
However, it is easier to recognize the correlation time  with the single exponential form, namely $\tau$.

We study size effects from 27 to 3375 atoms in the supercell. 
Convergence is achieved to 8\% for $216$ atoms and 3\% for $1728$ atoms. We use 1728 atoms for the production runs.

\section{Results}
We have chosen a set of $(V,T)$ conditions to sample the thermodynamic region that 
is relevant to the outer core (Table \ref{tab:viscosity}).
Comparing our pressures with existing data from experiments, first-principles molecular dynamics (FPMD) \cite{Alfe2000, Li2021} and embedded atom model molecular dynamics (EAMMD) \cite{Desgranges2007}, we find overall consistent results. The largest pressure deviation from  
experimental value occurs at condition $T= 6000, 7000$ \si{K}, $\rho=13.3$ \si{g/cm^3} and overestimates by about 18 \si{GPa}.  Due to the larger size of our system, pressure fluctuations are minimized, enhancing the overall stability and accuracy of the simulations. The integration of machine learning enables us to obtain a more detailed and extensive sampling of the liquid core region.

\begin{figure*}[htpb]
    \centering
    \includegraphics[width=0.95\textwidth]{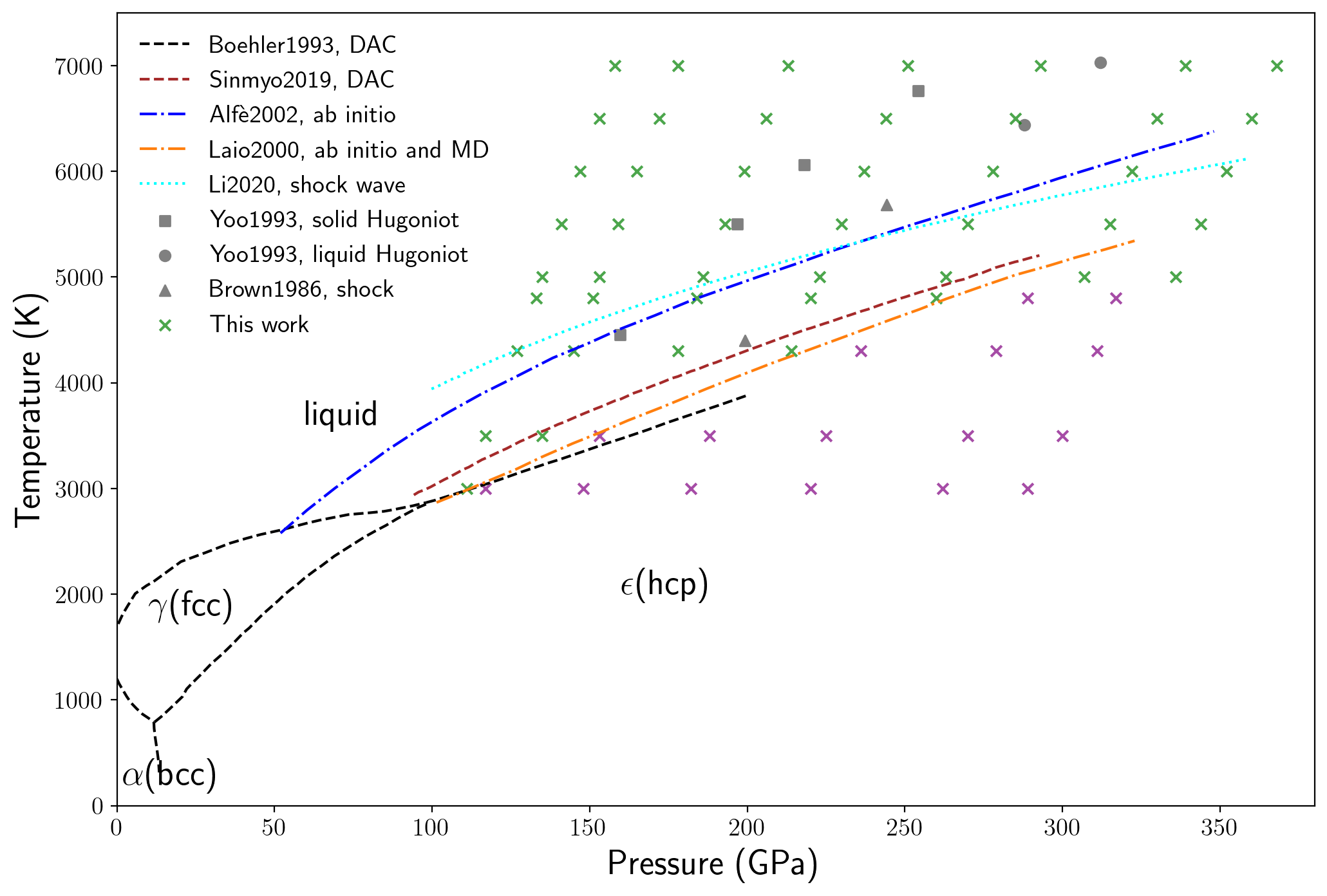}
    \caption{
    Conditions of our simulations, which were performed in cubic boxes at fixed volume, in $NVT$ ensemble. The phase diagrams from Refs. \cite{Boehler1986,Boehler1990,Boehler1993,Alfe2002,Laio2000, Li2020} are shown. The green crosses represent liquid and the purple  
    symbols show crystallization.} 
 \label{fig:phasediagram}
\end{figure*}

Structural and dynamical analyses, such as the PDF (pair distribution function) and mean squared displacement (MSD), are commonly used to distinguish between the liquid and solid states.
Alternatively, one can also use viscosity as an indicator. Examination of the asymptotic behavior of the running viscosity Eq. \ref{eq:gk}, 
can give a clue of a solid phase or a liquid phase (see purple vs green crosses in Figure \ref{fig:phasediagram}).
Our systems supercool before crystallizing, so all of our viscosities are for the pure liquid phase, including supercooled liquids.  

Computing a reliable melting line requires methods such as the two-phase method \cite{Yuan2023, Wu2024}. We have not determined the thermodynamic melting point, since a number of careful computational studies of the melting curve already exist \cite{Alfe2002} as well as experimental studies \cite{Li2020,Boehler1986,Boehler1990,Boehler1993,Sinmyo2019}.

\subsection{Diffusion coefficients}
We computed diffusivity $D$ using the \textit{compute msd} command from LAMMPS and performing a linear fit of the $\textrm{MSD} = 6 D t$. 
 The uncertainty in the diffusion coefficient in our simulation is negligible and thus omitted.
 Diffusion coefficient from DPMD agrees with FPMD within the reported uncertainty. For example, at $T=7000$ \si{K}, $\rho=10.7$ \si{g/cm^3}, 14.0 from DPMD is comparable to  FPMD's 
13.0$\pm 1.3$ in unit of \si{nm^2/s} (see Table \ref{tab:viscosity}.
Even within first-principles results, a large discrepancy remains.
For example, at T=5000 K, 10.7 \si{g/cm^3}, Alf\`e et. al. \cite{Alfe2000}
gives  $7.0 \pm 0.7$ but Li et. al. \cite{Li2021} gives $5.42 \pm 0.49$.
Our diffusivity value $6.6$ \si{nm^2/s} falls within the range reported by first-principles results. 

\begin{figure}[htpb]
    \centering
    \includegraphics[width=0.9\textwidth]{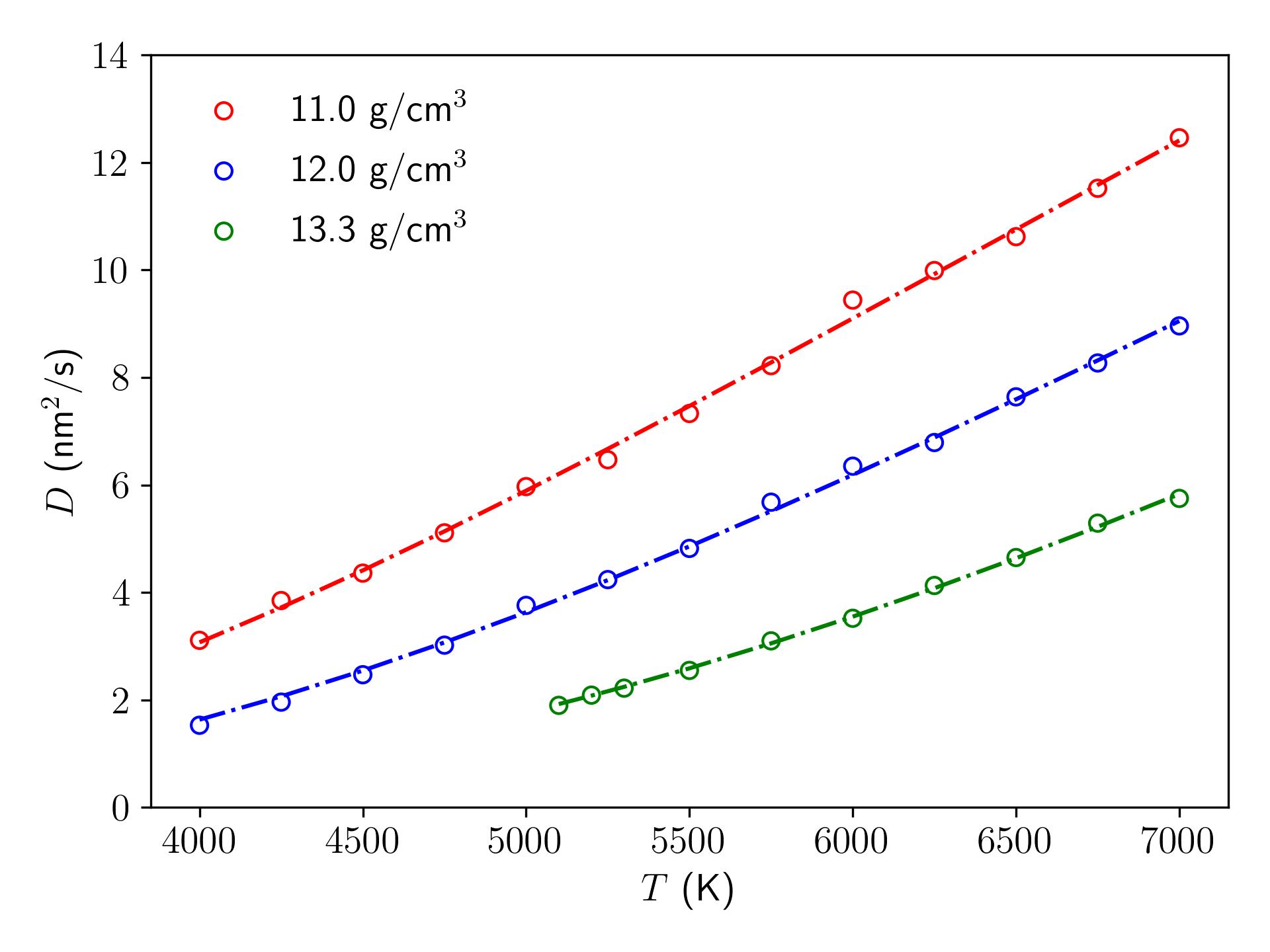}
    \caption{The diffusion coefficients at different temperatures for fixed densities of $\rho_1 = 11.0, \rho_2 = 12.0$, and $\rho_3 = 13.3$ in the unit of \si{g/cm^3}. 
    Fitting to the Arrhenius formula $D(T)=D_0 e^{-E_a / k_B T}$ yields
    $D_0=79.837$  \si{nm/s^2}, $E_a =1.123 $ \si{\eV} for $\rho_1$ (red), $D_0=88.683$ \si{nm/s^2}, $E_a = 1.377 $ \si{\eV} for $\rho_2$ (blue), and $D_0=114.407$ \si{nm/s^2}, $E_a = 1.796 $ \si{\eV} for $\rho_3$ (green).
    }
    \label{fig:diffusion_coeff}

\end{figure}

We test whether the diffusivity (Figure \ref{fig:diffusion_coeff}) and viscosity (Figure \ref{fig:viscosity}) vary with temperature at constant volume according to the Arrhenius relationship at densities, $\rho_1 = 11.0$ \si{g/cm^3}, $\rho_2 = 12.0$ \si{g/cm^3}, and $\rho_3 = 13.3$ \si{g/cm^3}. We find that the Arrhenius formula fits well for both quantities (see parameters in the captions). 

\subsection{Viscosity}

\begin{figure}[htpb]
    \centering
    \includegraphics[width=0.95\textwidth]{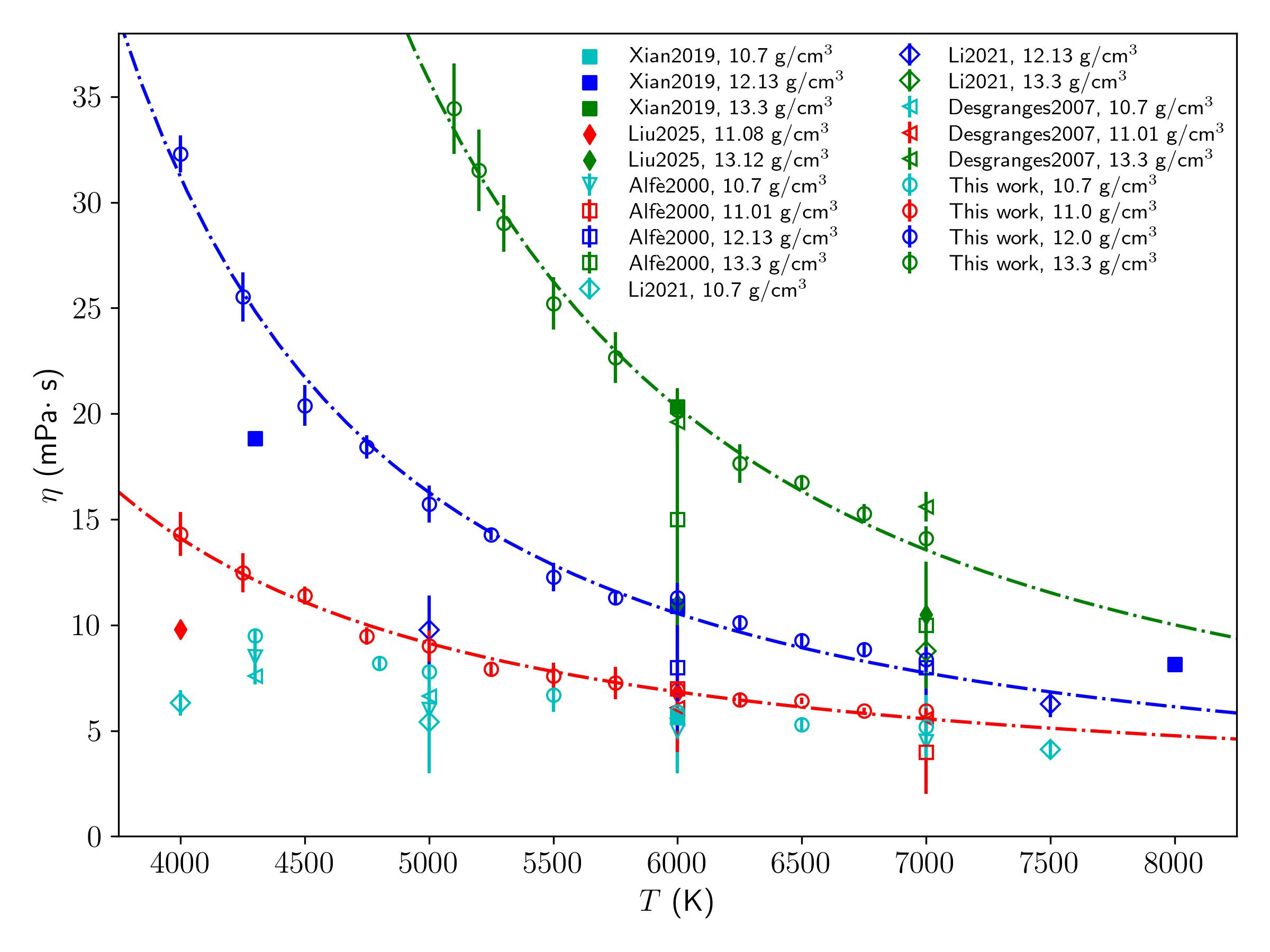}
    \caption{
    Data points from Xian et. al. \cite{Xian2019} (square), and Liu et. al. \cite{Cong2025} (thin diamond) without error estimation. 
    FPMD data from Alf\`e et. al. \cite{Alfe2000} (empty square), Li et. al. \cite{Li2021} (diamond),
    and EAMMD data from Desgranges and Delhommelle \cite{Desgranges2007} (triangle) as well as DPMD data from this work (circle) are included with error bars. 
    The viscosity data from DPMD is fitted to the Arrhenius formula (dashed line). Arrhenius parameters for viscosity, $\eta_0=0.748$ \si{mPa.s}, $E_a =1.608 $  \si{\eV} for $\rho_1$ (red), $\eta_0=1.121$ \si{mPa.s}, $E_a = 1.207 $  \si{\eV} for $\rho_2$ (blue),
    and $\eta_0=1.087$ \si{mPa.s}, $E_a = 1.513 $  \si{\eV} for $\rho_3$ (green). 
    }
 \label{fig:viscosity}
\end{figure}

We have reduced the uncertainty in viscosity from previous computations. For most conditions, the uncertainty remains below 0.8 \si{mPa.s}. 
In contrast, uncertainties derived from FPMD 
are consistently higher (see error bars in Fig. \ref{fig:viscosity}), typically exceeding 2.0 \si{mPa.s} \cite{Alfe2000}. 
For these three densities, to aid the comparison we thus use the same color (red for $\rho_1$, blue for $\rho_2$, and green for $\rho_3$) for similar densities from previously reported literature.
For convenience, we also associate the density $10.7$ \si{g/cm^3} with color cyan. 

For high density $\rho_3$, at T=6000 K, our predicted viscosity (about 20 \si{mPa.s}) agrees well with  first-principles results of Xian et. al. \cite{Xian2019} and EAMMD results of Desgranges and Delhommelle \cite{Desgranges2007}.
Alf\`e et. al. gives a rather lower viscosity about 15 \si{mPa.s} but with uncertain of 5 \si{mPa.s}, which could be deemed to match other results. At T=7000 K, the lowest of $8.77 \pm 1.72$  is given in Li et. al. \cite{Li2021}, while the highest of $15.6 \pm 0.7$ \si{mPa.s} is given in Ref. \cite{Desgranges2007}. Our prediction gives $14.1 \pm 0.6$ \si{mPa.s} while Alf\`e et. al. gives $8\pm 3$ \si{mPa.s}. 

For intermediate density $\rho_2$, large discrepancy occurs for T below 5000 K. Li et. al. gives $9.76 \pm 1.64$ \si{mPa.s} at $T=5000$ K,
which is much lower than our value of $15.7 \pm 0.9$ \si{mPa.s}. Similar trend occurs for Xian et. al. at lower $T= 4300$ K. Oddly, at $T=8000 $ K,  Xian et. al. gives a larger viscosity than our fitted line, and thus leads to a gentle slope in temperature.

For low density $\rho_1$, most predictions align with our fitted curve with the exception of Liu and Cohen \cite{Cong2025} at $T=4000$ K. The discrepancy reaches about 4 \si{mPa.s}. For $10.7$ \si{g/cm^3}, 
though the viscosity is already low, differences remain among existing literature. For $T=6000$ K, all data agree pretty well with each other. While for lower temperatures, Li et. al. gives evidently lower viscosity compared to other works.

Overall, our results agree well with previous viscosity estimates obtained from molecular dynamics simulations, whether using first-principles forces \cite{deWijs1998, Alfe2000, Pozzo2013, Li2021, Li2022_PEPI, Zhu2022} or empirical force fields \cite{Desgranges2007, Xian2019}. However, discrepancies arise under certain conditions, highlighting the sensitivity of viscosity calculations to methodological and force-field choices. 

\subsection{Einstein-Stokes relation}

\begin{figure}[h]
    \centering
    \includegraphics[width=0.9\textwidth]{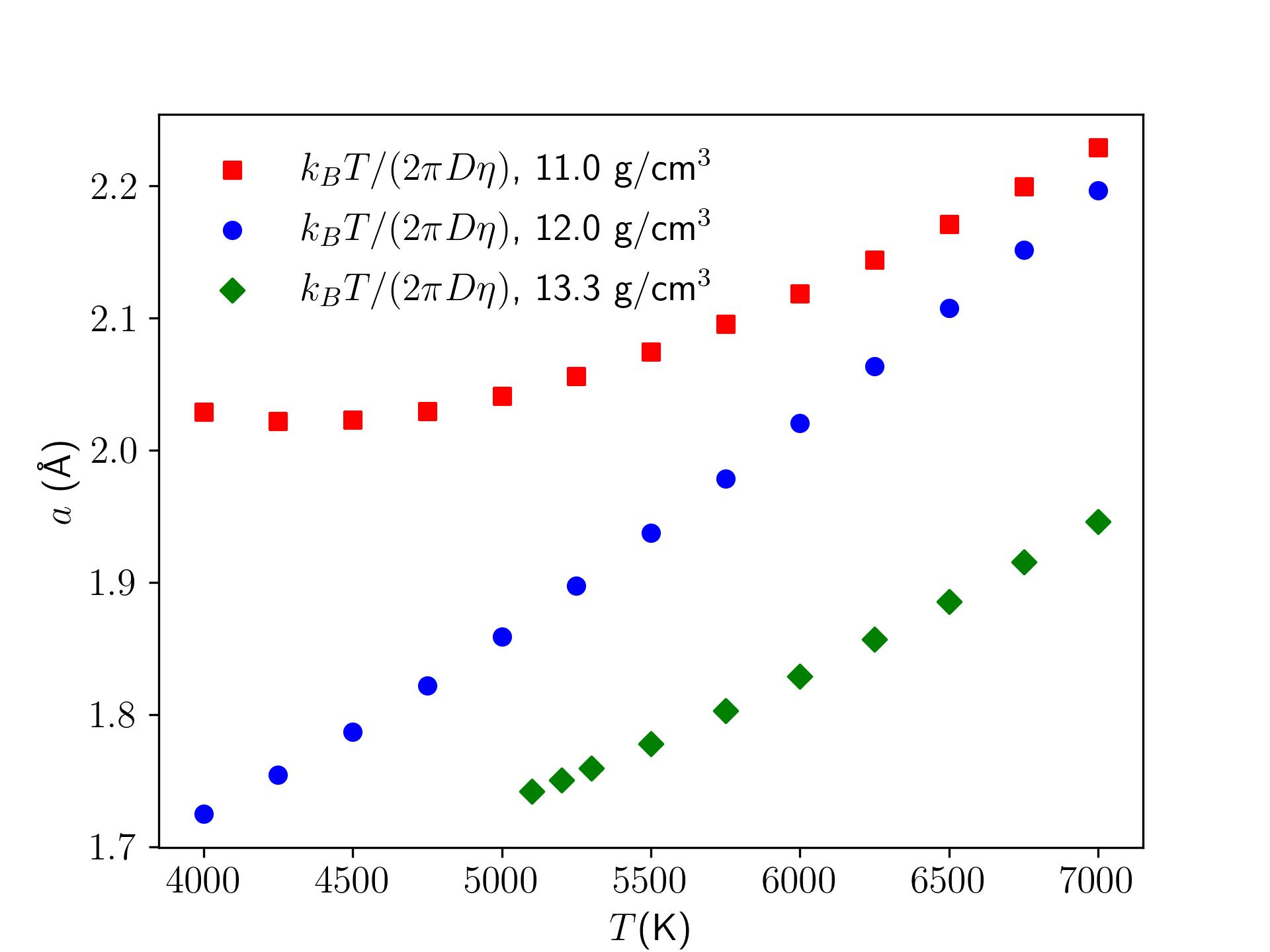}
    \caption{
    The Einstein-Stokes parameter $a=k_B T/(2\pi D\eta)$ is not constant, indicating failure of the Einstein-Stokes relation for iron.
    }
    \label{fig:einstein-stokes}
\end{figure}

We check the widely-used Einstein-Stokes relation between diffusivity and viscosity to see how well it holds. For the Einstein-Stokes relation to hold at a given density, the quantity $a=k_B T/(2\pi D\eta)$ should stay constant. We find that $a$ is not constant, and thus the Einstein-Stokes relation does not well fit our computational data. (Fig. \ref{fig:einstein-stokes}). This indicates that universal assumption of Einstein-Stokes relation for varying conditions is not an accurate assumption.

\subsection{Viscosity map}
\begin{figure}[htpb!]
    \centering
\includegraphics[width=0.90\textwidth]{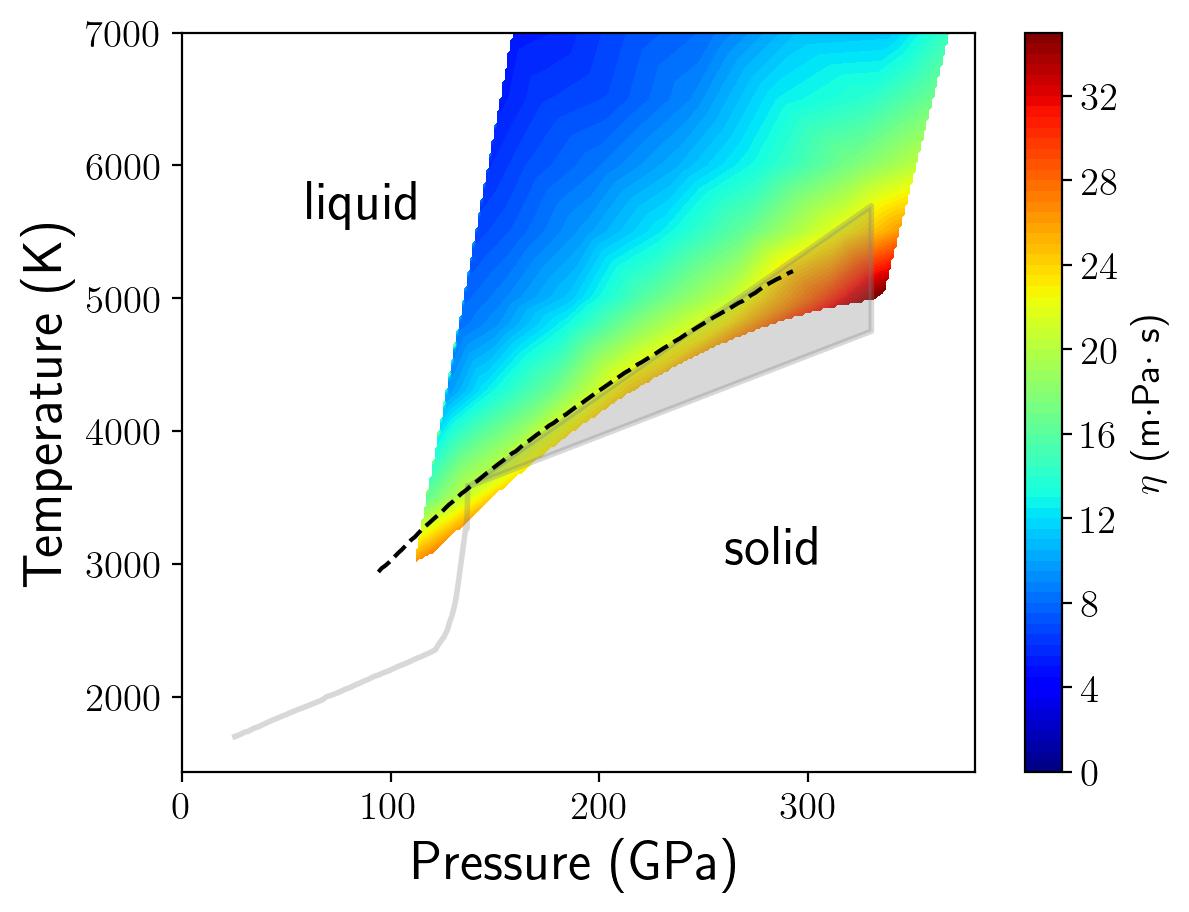}
\caption{Viscosity map under Earth's core conditions.
Gray lines are inferred temperature profile \cite{Boehler1993,Alfe2007,Stacey2008} and the black line is a melting curve from diamond anvil cell study \cite{Sinmyo2019}. 
}
 \label{fig:viscositymap}
\end{figure}

We generate a viscosity map by interpolating the viscosity data from Table \ref{tab:viscosity} onto a finer grid in the simulated thermodynamic region shown in Fig.(\ref{fig:viscositymap}). From this map, it can be easily seen  that the viscosity increases with increasing pressure and/or decreasing temperature. 
This map allows for easy determination of pressure-dependent (or depth-dependent) viscosity for a given geotherm. 

The resultant viscosity is on the order of tens of \si{mPas}. Such a low viscosity indicates that compared to Coriolis force and the Lorentz force,
viscous forces are negligible. The low viscosity of the outer core suggests a small Ekman number, which favors the inviscous fluid with a small-circulation turbulent 
convection picture \cite{Glatzmaiers1995}. Overall, the viscosity has a weak depth dependence (about a factor of 2), which reassures the usual treatment with non-depth dependent viscosity profile.

\section{Conclusions}
We have developed an electronic temperature dependent machine learning potential for irons under core conditions, which accurately reproduce DFT energies, forces and virials.  Utilized such machine learning potential in nanosecond scale molecular dynamics simulations allow us to simulate liquid iron under the conditions of Earth's core.
 We apply Green-Kubo formula to determine viscosity and diffusion coefficient of liquid iron. Subject to the pseudopotential and density functional approximation and 
 fitting errors in the machine learning, our simulations have achieved convergence with respect to simulation time and system size. 

By constructing a viscosity map for the liquid core, we can derive a depth-dependent viscosity profile tailored to geodynamo simulations for a given geotherm. This approach can be extended to generate analogous maps for other material properties of interest under extreme conditions.

Compared to empirical potentials, machine-learning potentials offer a more reliable description across a broader thermodynamic space. This machine-learning approach holds significant promise for investigating material properties under extreme conditions, particularly for problems requiring extended simulation times or large system sizes.

Future work of investigation of viscosity and other transport properties using the non-equilibrium molecular dynamics is underway.

\section{Acknowledgment}
This work is supported by the National Natural Science Foundation of China grant  12104230 and the US National Science Foundation CSEDI grant EAR-1901813 and the Carnegie Institution for Science. We gratefully acknowledge supercomputer support from the Resnick High Performance Computing Center. The authors gratefully acknowledge the Gauss Centre for Supercomputing e.V. (\href{http://www.gauss-centre.eu/}{www.gauss-centre.eu}) for funding this project by providing computing time on the GCS Supercomputer SuperMUC-NG at Leibniz Supercomputing Centre (\href{http://www.lrz.de/}{www.lrz.de}). The authors would like to express their gratitude to Qiyu Zeng, Han Wang and Linfeng Zhang for their valuable suggestions regarding the DPGEN settings. The authors are also thankful for the comments on the manuscript by Peter Driscoll.

\setlength{\tabcolsep}{1pt}
\renewcommand{\arraystretch}{0.5}
\begin{table}[htbp]

    \caption{
    Pressure, diffusion coefficient, and viscosity comparison from experiments and molecular dynamics with forces of first-principles (Alf\`e FP \cite{Alfe2000} and Li FP\cite{Li2021}), embedded atom model (EAM) \cite{Desgranges2007},
     and deep potential (DP) under Earth's core conditions.  Error estimates are in the parenthesis.
     \footnote{FP data in the table for $12.0$ \si{g/cm^3} are quoted for $12.13$ \si{g/cm^3}}
}
    \label{tab:viscosity}
    \scalebox{0.90}{
    \begin{tabular}{cc|cccc|cccc|cccc}
    \hline
    
    & & \multicolumn{4}{c}{$P$(\si{GPa})} & 
    \multicolumn{4}{c}{$D$ (\si{nm^2/s} ) } & 
    \multicolumn{4}{c}{$\eta$(\si{mPa.s})}\\
    \hline
    $T$(\si{K}) & $\rho$(\si{g/cm^3}) &  Expt. & Alf\`e FP &  EAM & DP  &  Alf\`e FP &   Li FP&  EAM & DP  & Alf\`e  FP & Li FP&  EAM  & DP \\
    \hline
   4300	& 10.7 & 135 & 132       &  135  & 127 & 5.2(0.2)     &           &  &  4.6 & 8.5(1.0)   &            &  7.6(0.4)  & 9.5(0.3)  \\
    4300	& 11.0 &     &   	 &       & 145 &              &           &  &  3.9 &            &            &            & 12.4(0.1) \\
    4300	& 11.5 &     &   	 &       & 178 &              &           &  &  2.9 &            &            &            & 16.8(0.2) \\
    4300	& 12.0 &     &   	 &       & 214 &              &           &  &  2.1 &            &            &            & 24.0(0.7) \\
    4300	& 12.5 &     &   	 &       & 236 &              &           &  &      &            &            &            &           \\
    4300	& 13.0 &     &   	 &       & 279 &              &           &  &      &            &            &            &           \\
    4300	& 13.3 &     &   	 &       & 311 &              &           &  &      &            &            &            &           \\
    4800	& 10.7 &     &   	 &       & 133 &              &           &  &  6.3 &            &            &            &  8.2(0.3) \\
    4800	& 11.0 &     &   	 &       & 151 &              &           &  &  5.4 &            &            &            &  9.4(0.2) \\
    4800	& 11.5 &     &   	 &       & 184 &              &           &  &  4.2 &            &            &            & 12.9(0.2) \\
    4800	& 12.0 &     &   	 &       & 220 &              &           &  &  3.2 &            &            &            & 18.1(0.2) \\
    4800	& 12.5 &     &   	 &       & 260 &              &           &  &  2.4 &            &            &            & 22.9(0.2) \\
    4800	& 13.0 &     &   	 &       & 289 &              &           &  &      &            &            &            &           \\
    4800	& 13.3 &     &   	 &       & 317 &              &           &  &      &            &            &            &           \\
    5000	& 10.7 & 145 & 140       &  141  & 135 & 7.0(0.7)     & 5.42(0.49)&  &  6.6 & 6.0(0.3)   & 5.6(0.2)   &  6.7(0.3)  &  7.8(0.2) \\
    5000	& 11.0 &     &   	 &       & 153 &              &           &  &  5.8 &            &            &            &  8.8(0.1) \\
    5000	& 11.5 &     &   	 &       & 186 &              &           &  &  4.6 &            &            &            & 11.2(0.6) \\
    5000	& 12.0 &     &   	 &       & 223 &              &  3.0(0.1) &  &  3.6 &            & 9.76(1.64) &            & 16.0(0.2) \\
    5000	& 12.5 &     &   	 &       & 263 &              &           &  &  2.8 &            &            &            & 20.5(0.2) \\
    5000	& 13.0 &     &   	 &       & 307 &              &           &  &  2.1 &            &            &            & 29.2(0.4) \\
    5000	& 13.3 &     &           &       & 336 &              &           &  &  1.8 &            &            &            & 35.9(0.6) \\
    5500	& 10.7 &     &   	 &       & 141 &              &           &  &  8.3 &            &            &            &  6.7(0.8) \\
    5500	& 11.0 &     &   	 &       & 159 &              &           &  &  7.6 &            &            &            &  7.6(0.5) \\
    5500	& 11.5 &     &   	 &       & 193 &              &           &  &  6.1 &            &            &            &  9.7(0.2) \\
    5500	& 12.0 &     &   	 &       & 230 &              &           &  &  5.0 &            &            &            & 12.5(0.4) \\
    5500	& 12.5 &     &   	 &       & 270 &              &           &  &  3.7 &            &            &            & 16.5(0.1) \\
    5500	& 13.0 &     &   	 &       & 315 &              &           &  &  3.0 &            &            &            & 22.3(0.4) \\
    5500	& 13.3 &     &   	 &       & 344 &              &           &  &  2.6 &            &            &            & 26.3(0.5) \\
    6000	& 10.7 & 155 & 151	 &  149  & 147 & 10.0(1.0)    &           &  & 10.3 &  5.0(2.0)  &          &   5.6(0.3) & 5.9(0.2)  \\
    6000	& 11.0 & 170 & 170 	 &  166  & 165 &  9.0(0.9)    &           &  &  9.0 &  7.0(3.0)  &           &  6.1(0.3)  &  7.0(0.4) \\
    6000	& 11.5 &     &   	 &       & 199 &              &           &  &  7.3 &            &            &            &  8.4(0.3) \\
    6000	& 12.0 & 240 & 251 	 &       & 237 &  6.0(0.6)    &           &  &  6.1 & 8.0(3.0)   & 	      &            & 10.4(0.2) \\
    6000	& 12.5 &     &    	 &       & 278 &              &           &  &  5.1 &            &            &            & 13.4(0.1) \\
    6000	& 13.0 &     &    	 &       & 322 &              &           &  &  4.1 &            &            &            & 17.5(0.2) \\
    6000	& 13.3 & 335 & 360 	 &  332  & 352 &  5.0(0.5)    &           &  &  3.6 & 15.0(5.0)  &            & 19.6(1.0)  & 20.2(0.2) \\
    6500	& 10.7 &     &    	 &       & 153 &              &           &  & 12.2 &            &            &            &  5.3(0.3) \\
    6500	& 11.0 &     &   	 &       & 172 &              &           &  & 11.1 &            &            &            &  6.4(0.7) \\
    6500	& 11.5 &     &   	 &       & 206 &              &           &  &  9.0 &            &            &            &  7.1(0.8) \\
    6500	& 12.0 &     &    	 &       & 244 &              &           &  &  7.4 &            &            &            &  9.0(0.3) \\
    6500	& 12.5 &     &     	 &       & 285 &              &           &  &  6.3 &            &            &            & 12.5(0.3) \\
    6500	& 13.0 &     &   	 &       & 330 &              &           &  &  5.1 &            &            &            & 14.2(0.4) \\
    6500	& 13.3 &     &   	 &       & 360 &              &           &  &  4.8 &            &            &            & 16.8(0.3) \\
    7000	& 10.7 &     & 161       &       & 158 & 13.0(1.3)    &           &  & 14.0 &  4.5(2.0)  &            &            &  5.2(1.5) \\
    7000	& 11.0 &     & 181       &  174  & 178 & 11.0(1.1)    &           &  & 12.5 &  4.0(2.0)  & 5.6(0.3)   &            & 5.5(0.7)  \\
    7000	& 11.5 &     &   	 &       & 213 &              &           &  & 10.8 &            &            &            &  7.0(0.2) \\
    7000	& 12.0 & 250 & 264 	 &       & 251 &  9.0(0.9)    &           &  &  9.3 &  8.0(3.0)  &            &            &  8.1(0.5) \\
    7000	& 12.5 &     &   	 &       & 293 &              &           &  &  7.5 &            &            &            & 10.9(1.6) \\
    7000	& 13.0 &     &   	 &       & 339 &              &           &  &  6.6 &            &            &            & 11.2(0.5) \\
    7000	& 13.3 & 350 & 375 	 &       & 368 &  6.0(0.6)    &  5.1(0.2) &  &  5.8 & 10.0(3.0)  & 8.77(1.72) & 15.6(0.7)  & 14.7(0.6) \\
    \hline
    \end{tabular}
    }
\end{table}


\bibliography{ref} 

\end{document}